\documentclass[final,1p,times]{elsarticle}
\usepackage[T1]{fontenc}
\usepackage{amsmath,amssymb,bm,booktabs,graphicx,tabularx}
\usepackage{xcolor}
\usepackage[colorlinks=true,linkcolor=blue!45!black,citecolor=blue!45!black,urlcolor=blue!45!black]{hyperref}
\usepackage{microtype}
\graphicspath{{figures/}}
\newcommand{\mevA}{\mathrm{meV}\,\text{\AA}^{-1}}
\newcommand{\mevat}{\mathrm{meV}\,\mathrm{atom}^{-1}}
\newcommand{\mjm}{\mathrm{mJ}\,\mathrm{m}^{-2}}

\newcommand{\RMS}{\operatorname{RMS}}

\journal{Manuscript draft}
\makeatletter
\patchcmd{\MaketitleBox}{\elsaddress\par\vskip36pt}{\elsaddress\par\vskip12pt}{}{}
\def\ps@pprintTitle{\let\@oddhead\@empty\let\@evenhead\@empty
  \def\@oddfoot{\footnotesize\itshape Working manuscript\hfill26 October 2026}
  \let\@evenfoot\@oddfoot}
\makeatother
\begin{document}
\begin{frontmatter}
\title{Retention-constrained fine-tuning of a pre-trained aluminium MACE potential for non-equilibrium defect dynamics and plasticity}
\author{\texorpdfstring{Yue Wu$^{1,2}$, Yuan Tian$^2$, Deng Pan$^{1,*}$ and Mingwei Chen$^{2,*}$}{Yue Wu, Yuan Tian, Deng Pan and Mingwei Chen}}
\address{$^1$ Materials Genome Institute, Shanghai University, Shanghai 200444, China}
\address{$^2$ State Key Laboratory of Quantum Functional Materials, Department of Materials Science and Engineering, and Institute of Innovative Materials, Southern University of Science and Technology, Shenzhen, 518055, China}
\begin{abstract}
Adapting a pre-trained machine-learning interatomic potential for non-equilibrium defect dynamics and plasticity requires improving accuracy on far-from-equilibrium configurations, such as loaded grain boundaries, planar-slip and twinning intermediates, without degrading the near-equilibrium elastic and vibrational response that governs the mechanical fields surrounding those defects. Here we achieve this balance for the aluminium MACE potential QP17 by integrating conditional readout uncertainty, iterative full-parameter updates and protocol-compatible relative-energy anchoring, and we audit retention on the lattice constant, independent elastic constants and harmonic phonons that set the bulk mechanical response. By evaluating sensitivities over 144 linear readout weights while updating the covariance within each acquisition batch and evolving the upstream neural representation between rounds, our framework targets weakly constrained defect-sensitive parameter directions efficiently. After four active-learning rounds (32 acquired configurations and 400 gradient updates), force-uncertainty acquisition reduces the mean development force error from 22.43 to 21.52 $\mevA$, outperforming random sampling (22.70 $\mevA$), geometry-diversity selection (22.46 $\mevA$) and uniform training across all 104 candidate labels under the same update budget (22.71 $\mevA$). Crucially, all three force-uncertainty runs satisfy a strict suite of bulk-mechanical retention criteria: historical relative-energy error decreases from 0.375 to 0.207 $\mevat$, while harmonic phonon root-mean-square (RMS) drift remains below the 0.05 THz threshold. Matched objective ablations reveal that relative-energy anchoring suppresses harmonic drift, consistent with constraints on sampled near-equilibrium potential-energy curvature. Out-of-sample evaluations across planar faults, twin boundaries and grain boundaries demonstrate transferable gains in defect energies and deviatoric stresses, the latter improving even though virial stresses are never fitted, indicating that the adapted potential improves mechanical prediction, not only energetic prediction. Finally, a closed-loop molecular dynamics and density functional theory (MD-DFT) pilot confirms that automated bulk-mechanical acceptance gating intercepts updates that violate historical elastic-group energy or vibrational tolerances. Together, these components establish a retention-constrained fine-tuning protocol for adapting pre-trained MACE potentials to plastic-deformation observables, in which gains in defect energies and deviatoric stresses are retained only when the bulk mechanical response, including lattice constant, elastic constants and harmonic phonons, remains within prescribed tolerances.
\end{abstract}
\end{frontmatter}
\section{Introduction}
Machine-learning interatomic potentials (MLIPs) have enabled atomistic simulations of crystalline defects at length and time scales far beyond the reach of direct electronic-structure methods \cite{behler2007,bartok2018,zhang2024iron}. Once a baseline potential has been trained, practical deployment frequently demands selective refinement: enhancing predictive fidelity across application-specific defect topologies while preserving the equilibrium, elastic and vibrational responses already captured by the model \cite{wen2021}. The intended deployment regime is non-equilibrium: plastic deformation proceeds through the nucleation and motion of dislocations, the formation and migration of stacking faults and twin boundaries, and the loading of grain boundaries, all of which require configurations far from mechanical equilibrium \cite{vanswygenhoven2004,lu2000}. Aluminium exemplifies this tension. Accurately modelling distorted grain boundaries, planar stacking faults and twinning intermediates requires sampling far-from-equilibrium local environments, yet the mechanical stability and long-range stress fields of these defects remain governed by the underlying bulk elastic moduli and harmonic lattice dynamics \cite{mishin1999}. A potential intended for deformation simulations must therefore predict defect energies, deviatoric stresses and slip-path energetics accurately in these non-equilibrium environments, while the near-equilibrium response that sets the mechanical background is preserved. Effective fine-tuning must therefore resolve a plasticity--stability trade-off, maximizing target-domain gains on non-equilibrium defect observables while strictly bounding drift in the bulk mechanical response.

Active learning offers a principled route to allocate first-principles reference calculations toward maximally informative atomic configurations \cite{zhang2020dpgen,vandermause2020}. Early linear and kernel-based potentials leveraged parameter covariance or D-optimality extrapolation grades to trigger data acquisition \cite{podryabinkin2017,jinnouchi2019}, while recent prediction-rigidity formalisms directly link local uncertainty to the constraints imposed by the training distribution \cite{chong2025}. In parallel, Callsen, Lee and Chou \cite{callsen2026} demonstrated that combining Bayesian error estimation with dynamic force-constant training allows trajectory-driven uncertainty excursions to guide both configuration selection and convergence assessment. Extending this closed-loop philosophy to deep equivariant neural potentials \cite{batzner2022}, however, introduces a second challenge: regulating the structural collateral of gradient-based fine-tuning. Higher-order message-passing architectures such as MACE provide expressive many-body representations of potential-energy surfaces \cite{batatia2022,batatia2025foundation}, and recent work has highlighted both the promise and fragility of fine-tuning pre-trained foundation potentials for downstream materials properties, phonon dispersions and phase transitions \cite{radova2025,grandel2026,deng2025}. When adapting such models to high-strain defects, informative data selection alone cannot prevent unconstrained parameter updates from distorting near-equilibrium curvature that governs the mechanical response of the defects being studied. Complementing active acquisition with historical replay, protocol-compatible energy-landscape anchoring and explicit bulk-mechanical gating provides a structured mechanism to assimilate new defect physics without erasing the foundational bulk behaviour on which deformation simulations depend. This work can be read as a concrete instantiation of the physics-driven specialization route outlined by Wu et al. \cite{wu2026specialized}, which distinguishes broad chemical coverage from validated fidelity under demanding physical conditions and identifies targeted sampling, fine-tuning and physical constraints as the levers for closing that gap; we exercise all three for a single application context, non-equilibrium defect dynamics in aluminium, and supply the observable-specific retention evidence that the framework treats as the decisive test.

Here we formulate and validate a retention-constrained fine-tuning workflow for the pre-trained aluminium MACE potential QP17 that couples conditional readout uncertainty with relative-energy anchoring and multi-property retention checks. By operating on the linear readout layers to rank candidate structures and updating the precision matrix sequentially within each acquisition batch, our strategy suppresses sampling redundancy while permitting full-network representation learning between rounds. Through matched loss ablations, multi-round pool benchmarks against random, geometric and full-pool baselines, out-of-sample defect-archive and planar-slip evaluations, and a trajectory-driven MD--DFT pilot, we demonstrate how targeted data allocation and curvature-anchoring regularization define a reliable operating window for defect adaptation under explicit bulk-mechanical retention.

\section{Methodology}
\subsection{Workflow architecture and experimental design}

\begin{figure}[tbp]
\centering\includegraphics[width=\textwidth]{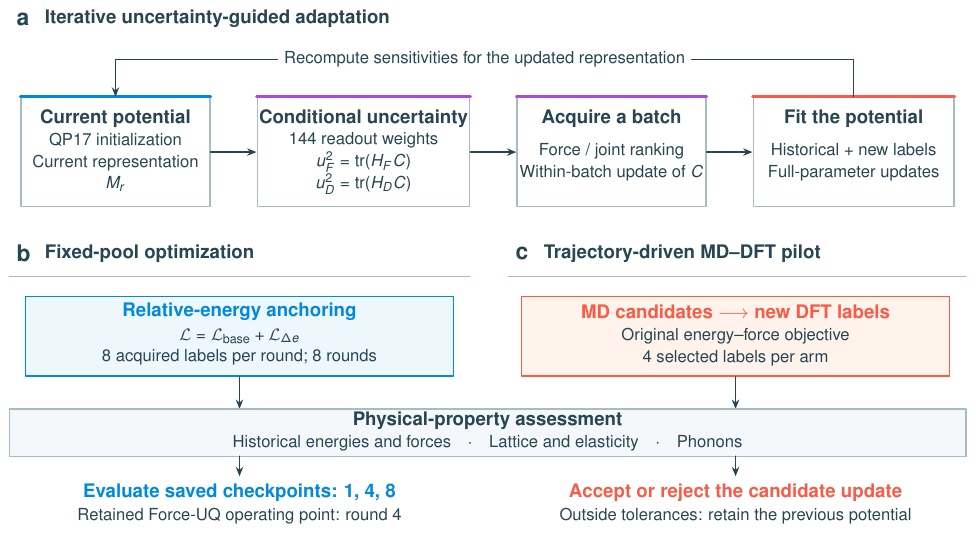}
\caption{\textbf{Retention-constrained fine-tuning workflow for non-equilibrium defect dynamics.} \textbf{a}, Iterative active-learning loop: conditional sensitivities over the 144 linear readout weights rank far-from-equilibrium candidate configurations, sequential within-batch precision updates suppress redundant selections, and full-parameter gradient updates evolve the neural representation $M_r$ for subsequent rounds. \textbf{b}, Fixed-pool benchmark protocol combining historical replay with pairwise relative-energy anchoring ($\mathcal{L} = \mathcal{L}_{\mathrm{base}} + \mathcal{L}_{\Delta e}$, eight acquired labels and 100 updates per round across eight rounds), with comprehensive physical audits at rounds 1, 4 and 8 identifying round 4 as the latest retention-compliant Force-UQ checkpoint among these evaluations. \textbf{c}, Trajectory-driven MD--DFT pilot acquiring newly labelled first-principles configurations online and enforcing automated bulk-mechanical acceptance or rejection to protect subsequent dynamics.}
\label{fig:workflow}
\end{figure}

Our adaptation framework integrates three functional stages: sensitivity-driven batch acquisition, relative-energy-anchored optimization and physical-property acceptance gating (Fig.~\ref{fig:workflow}). All fine-tuning experiments initialize from the pre-trained aluminium MACE checkpoint QP17 and update all trainable network parameters. The fixed-pool benchmark operates on a curated dataset comprising 79 historical replay configurations and a candidate pool of 104 configurations spanning perturbed face-centred cubic (FCC) and hexagonal close-packed (HCP) bulk supercells, planar-slip and twinning intermediates, and mechanically loaded $\Sigma 3$ and $\Sigma 5$ grain boundaries. Target domain accuracy is tracked on a held-out development set of 24 bulk and defect configurations, while bulk-mechanical retention is audited independently on an eight-configuration historical holdout set excluded from replay. The audited quantities, lattice constant, independent elastic constants and harmonic phonons, are those that determine the long-range stress fields and mechanical stability of the defect configurations under study. To evaluate transfer across distinct defect environments outside the training and development sets, we also assemble a retrospective defect archive containing eight scored configurations and three reference states, screened via radial-descriptor overlap prior to model evaluation. Complete dataset provenance, first-principles protocols and label-eligibility masks are detailed in the Supporting Information (SI).

In the primary eight-round benchmark, we evaluate four sequential acquisition strategies---force uncertainty quantification (Force UQ), joint force--deviatoric-stress uncertainty (Joint UQ), geometry-diversity selection and uniform random sampling---across three paired random seeds. Each branch acquires eight previously unselected configurations per round and performs 100 gradient updates (totalling 800 updates over eight rounds). To isolate the role of newly acquired defect data from continued optimization and label volume, we include two reference controls under an identical 800-update schedule: a historical-only branch (\texttt{Old only}) trained exclusively on the 79 replay configurations across three seeds, and a single-seed full-pool baseline (\texttt{All 104}) trained uniformly on all 104 candidate labels from the first round. An auxiliary two-seed multi-round experiment using the unmodified energy--force loss provides a baseline for assessing how energetic anchoring modulates planar-slip adaptation. Prior to the multi-round comparisons, a matched $2 \times 2$ objective ablation is conducted across two eight-configuration training sets---a fixed candidate-pool subset and the newly labelled frames from the MD--DFT pilot (evaluated against six independent-trajectory audit frames)---to establish the common fine-tuning objective. All historical holdout, defect-archive and planar-slip benchmarks remain strictly external to objective selection and training.

\subsection{Conditional readout uncertainty and batch-aware acquisition}

Computing exact Bayesian posteriors over all parameters of an equivariant message-passing network is computationally intractable during iterative active learning. To achieve analytically exact within-batch information updates while preserving full-network adaptability between rounds, we formulate a conditional linear-readout uncertainty metric over the 144 linear readout weights of QP17 (comprising 128 weights in the first linear readout layer and 16 in the second). During acquisition scoring, upstream message-passing parameters are held fixed to extract readout Jacobians; following full-parameter gradient descent, these sensitivities are recomputed at the updated latent representation at the start of the next round (Fig.~\ref{fig:workflow}a).

For a candidate configuration $X$ containing $N$ atoms, let $j_e(X) \in \mathbb{R}^{144}$ denote the gradient of the per-atom energy with respect to the readout weights, and let $J_F(X) \in \mathbb{R}^{3N \times 144}$ and $J_D(X) \in \mathbb{R}^{5 \times 144}$ denote the corresponding Jacobians of the $3N$ Cartesian force components and the five orthonormal deviatoric-stress components, respectively. We define the normalized per-configuration force and deviatoric-stress sensitivity matrices as
\begin{equation}
H_F(X) = \frac{J_F^\top J_F}{3N}, \qquad H_D(X) = \frac{J_D^\top J_D}{5}. 
\label{eq:h}
\end{equation}

Conditioned on the cumulative set of acquired configurations $S$, the readout parameter covariance matrix $C$ is constructed as
\begin{equation}
C = \left[ G_{\mathrm{hist}} + \lambda I + \sum_{X \in S} \left( \frac{j_e j_e^\top}{s_E^2} + \frac{H_F}{s_F^2} \right) \right]^{-1}, 
\label{eq:cov}
\end{equation}

with fixed noise-scaling factors $s_E = 0.020\text{ eV atom}^{-1}$ and $s_F = 0.085\text{ eV \AA}^{-1}$. The historical prior precision matrix $G_{\mathrm{hist}} = \sum_{X \in \mathcal{H}} j_e(X) j_e(X)^\top / s_E^2$ is accumulated over $|\mathcal{H}| = 837$ deduplicated historical geometries and regularized by $\lambda = 10^{-6}\lambda_{\max}(G_{\mathrm{hist}})$. These 837 geometries constrain the prior readout sensitivity directions, while 79 qualified historical configurations supply the replay labels used for gradient updates.

Predictive acquisition variances for atomic forces and deviatoric stresses are evaluated via the trace projections
\begin{equation}
u_F^2(X) = \mathrm{tr}[H_F(X)C], \qquad u_D^2(X) = \mathrm{tr}[H_D(X)C]. 
\label{eq:uq}
\end{equation}

Within each eight-configuration acquisition round, Force UQ greedily selects the candidate maximizing $u_F^2(X)$, whereas Joint UQ maximizes the equal-weight mean of the ascending percentile ranks of $u_F^2(X)$ and $u_D^2(X)$ across the remaining pool. Crucially, because the expected energy--force information contribution $j_e j_e^\top / s_E^2 + H_F / s_F^2$ depends exclusively on atomic coordinates and current network derivatives rather than reference DFT labels, selecting a candidate immediately updates the precision matrix in Eq.~\eqref{eq:cov} before the next candidate in the batch is ranked. This sequential information update penalizes collinear sensitivity directions within the same round, reducing batch redundancy while keeping DFT labels strictly unseen until selection is finalized.

As a non-Bayesian structural baseline, geometry-diversity selection characterizes each configuration using standardized radial-environment descriptors (combining per-atom feature means and standard deviations with the logarithm of atomic volume) and iteratively selects the candidate maximizing the minimum Euclidean distance to the already acquired set following a random initial draw. Random acquisition samples candidates uniformly without replacement. All acquisition strategies feed into an identical downstream training objective.

\subsection{Protocol-compatible energy anchoring and optimization}

For an individual configuration, the base fitting loss $\mathcal{L}$ combines the squared per-atom energy residual (scaled by $s_E$) and the mean squared force-component residual (scaled by $s_F$), subject to protocol-specific eligibility masks. Each stochastic gradient step samples one configuration from the historical replay buffer and one from the cumulative acquired pool to minimize
\begin{equation}
\mathcal{L}_{\mathrm{base}} = (1 - w)\mathcal{L}_{\mathrm{old}} + w\mathcal{L}_{\mathrm{new}}, 
\label{eq:base}
\end{equation}

where $w$ controls the relative emphasis on newly acquired defect data. Because historical datasets compiled across multiple studies often employ distinct DFT $k$-point meshes or plane-wave cutoffs, constraining absolute total energies directly can introduce artificial offset conflicts. To anchor the curvature and relative energetics of the historical potential-energy surface while remaining invariant to group-specific reference offsets, we introduce a within-group pairwise relative-energy penalty:
\begin{equation}
\mathcal{L}_{\Delta e} = \frac{\left[(e_i - e_j) - (e_i^{\mathrm{DFT}} - e_j^{\mathrm{DFT}})\right]^2}{2s_{\Delta e}^2}, \qquad s_{\Delta e} = 0.001\text{ eV atom}^{-1}. 
\label{eq:pair}
\end{equation}

At each optimization step, an eligible historical protocol group (spanning bulk, strained and defect subsets) is drawn uniformly at random, and two configurations $(i, j)$ are sampled independently with replacement from within that group.

To identify a regularization regime among the tested settings, our $2 \times 2$ objective ablation crosses two new-data weightings---equal weighting ($w = 0.500$) and a cardinality-reduced weighting ($w = 2N_{\mathrm{batch}} / (N_{\mathrm{old}} + 2N_{\mathrm{batch}}) = 16/95 \approx 0.168$)---with the presence or absence of $\mathcal{L}_{\Delta e}$. Each of the four candidate objectives is trained for 100 updates on both the eight-frame pool subset and the eight-frame MD-derived training set. Among the formulations satisfying all physical-retention guardrails across both datasets, the setting yielding the lowest mean normalized force error on the development and MD-audit sets ($w = 0.500$ with $\mathcal{L}_{\Delta e}$; Section~\ref{sec:learner}) is selected as the standard learner. All runs utilize the Adam optimizer (learning rate $10^{-5}$, single precision), resetting optimizer momentum states at the start of each round while carrying forward network weights and the cumulative acquired pool. In the \texttt{Old only} control, both terms in Eq.~\eqref{eq:base} draw from the 79-frame replay buffer alongside $\mathcal{L}_{\Delta e}$. Because virial stress labels are omitted from the training loss across all branches, comparing Joint UQ against Force UQ specifically isolates the benefit of incorporating strain-derivative sensitivity during data acquisition.

\subsection{Physical-property acceptance guardrails and observable evaluation}

To prevent target-domain overfitting from degrading bulk mechanical response that governs defect stress fields, model checkpoints must satisfy a conjunctive suite of physical-retention criteria bounding historical replay force and relative-energy errors, group-centred historical energy errors, the equilibrium FCC lattice constant ($a_0$), independent elastic constants ($C_{11}, C_{12}, C_{44}$), bulk modulus ($B$) and harmonic phonon stability. Specifically, relative drift in $a_0$ is restricted to $\le 0.2\%$, elastic constant deviations are bounded by $|C-C_{\mathrm{QP17}}|\leq\max(0.05|C_{\mathrm{QP17}}|,2\text{ GPa})$ for each elastic quantity $C$, and harmonic phonon RMS drift and maximum branch shifts relative to QP17 are capped at $0.05\text{ THz}$ and $0.20\text{ THz}$, respectively (complete threshold definitions are tabulated in the SI). No composite weighting is permitted to trade a violation in one physical observable against gains in another. In the fixed-pool benchmark, full physical audits are executed at scheduled rounds 1, 4 and 8 to identify the latest retention-compliant operating point among the scheduled checkpoints, whereas in the closed-loop MD pilot these guardrails act as an automated gate governing whether an updated checkpoint replaces the active potential (Fig.~\ref{fig:workflow}b,c).

For physical observable evaluations, $a_0$ is obtained via structural energy minimization, and $C_{11}, C_{12}, C_{44}$ and $B$ are computed from central finite differences of stress at each model's equilibrium volume. Harmonic phonon dispersions are evaluated using Phonopy \cite{togo2023} at a standardized lattice constant of $4.045\text{ \AA}$ within a $5 \times 5 \times 5$ primitive supercell (125 atoms) using $\pm 0.01\text{ \AA}$ atomic displacements across an 81-point $\Gamma\!\text{--X--W--K--}\!\Gamma\!\text{--L}$ reciprocal-space path. Throughout this study, QP17 defines the baseline reference for quantifying functional drift in lattice, elastic and phonon properties, whereas DFT labels define reference predictive accuracy. Shaded bands and error bars represent minimum-to-maximum ranges across paired random seeds.

Because the present claim concerns plastic-deformation observables, competing planar-slip pathways are evaluated using a 48-atom constrained slab initialized in a precursor configuration bearing intrinsic shears on the first and third of three adjacent $(111)$ planes (denoted \texttt{101}). Route P2 completes slip on the central plane to nucleate a three-layer micro-twin (\texttt{101} $\rightarrow$ \texttt{111}), whereas route P3 reverses the outer shear to restore an intrinsic stacking fault (\texttt{101} $\rightarrow$ \texttt{100}). Each pathway is discretized into 11 prescribed in-plane shear coordinates ($q \in [0, 1]$) with structural relaxation restricted normal to the slip plane. Fault energies ($\Delta\gamma$) are normalized by interface area and referenced to each model's relaxed \texttt{101} precursor. Pathway accuracy is quantified via the RMSE across the three non-zero DFT reference coordinates ($q = 0.3, 0.5, 1.0$) per route, supplemented by single-point evaluations at fixed DFT reference geometries. For the retrospective defect archive, configurations retain their native DFT protocols; previously established $k$-mesh force sensitivities of 10--14 $\mevA$ for the $\Sigma 5$ and incoherent-twin families provide the numerical sensitivity scale when interpreting fine inter-method differences.

\subsection{Trajectory-driven MD sampling and DFT audit}

The closed-loop MD--DFT pilot tests online exploration and automated acceptance gating starting from QP17 across two periodic systems: a 24-atom bulk FCC supercell at $300\text{ K}$ and a 36-atom thin-twin supercell at $600\text{ K}$. Independent initial velocity seeds generate paired exploration and audit trajectories for each system, propagated for $2\text{ ps}$ ($1\text{ fs}$ timestep) under NVT Langevin dynamics in ASE \cite{larsen2017}. Following a $200\text{ fs}$ equilibration window, conditional force and deviatoric-stress uncertainties are evaluated every $50\text{ fs}$, where the frame-level force uncertainty is defined as the 90th percentile of per-atom RMS force uncertainties to prevent bulk atoms from diluting localized defect signals. Following the relative trajectory-average strategy of Callsen et al. \cite{callsen2026}, an uncertainty excursion is flagged whenever force or stress uncertainty exceeds three times its running mean over the preceding ten sampled frames. To guarantee active testing even in quiescent trajectories where no threefold excursion is triggered, a prescribed exploration quota selects two frames per system separated by at least $200\text{ fs}$---ranked by joint force--stress uncertainty in the Joint-UQ arm and drawn uniformly in the random arm.

Each branch is updated for 100 steps on its four newly labelled exploration frames under the original energy--force objective ($w = 0.500$). Six additional frames extracted from the independent audit trajectories are labelled with DFT and held out from fitting (four serving as recurring accuracy monitors and two providing initial baseline diagnostics). First-principles calculations for the 14 unique trajectory snapshots and two $k$-mesh convergence checks (16 single-point evaluations in total) are performed using VASP \cite{kresse1996} with the PBE exchange--correlation functional \cite{perdew1996} and projector augmented-wave (PAW) potentials \cite{blochl1994} at a $600\text{ eV}$ plane-wave cutoff and a $10^{-8}\text{ eV}$ electronic convergence criterion.

\section{Results and discussion}
\subsection{Relative-energy anchoring is required to retain harmonic response under non-equilibrium updates}
\label{sec:learner}

The central question for a plasticity-targeted potential is whether an accuracy gain on non-equilibrium configurations can be obtained without perturbing the near-equilibrium curvature that sets the elastic and vibrational background. We first isolate the respective contributions of uncertainty-guided configuration selection and relative-energy regularization through single-round component tests. Under the unmodified energy--force objective ($w = 0.500$, without $\mathcal{L}_{\Delta e}$), acquiring a single batch of eight candidate configurations via Force UQ (which exclusively selects mechanically loaded $\Sigma 5$ grain-boundary structures) reduces the development force RMSE from the QP17 baseline of $22.43~\mevA$ to $20.56~\mevA$ after 100 gradient updates. By contrast, acquiring eight configurations at random yields $22.39~\mevA$ under the identical update budget. At this initial stage, baseline drift remains minimal: historical replay force RMSE shifts slightly from $8.48$ to $8.56~\mevA$, accompanied by a modest phonon RMS drift of $0.0167\text{ THz}$. Conditional readout sensitivity thus extracts high-leverage defect information that random sampling misses.

When updates are driven by more strongly perturbed trajectory snapshots or extended across multiple rounds, however, fitting without relative-energy anchoring perturbs bulk lattice dynamics. Our matched $2 \times 2$ objective ablation on the eight MD-derived training configurations demonstrates how relative-energy anchoring resolves this vulnerability (Table~\ref{tab:ablation}). Training under the original objective ($w = 0.500$) reduces the six-frame independent audit force RMSE from $46.45~\mevA$ (QP17) to $41.36~\mevA$, but induces a phonon RMS drift of $0.0645\text{ THz}$, violating the $0.05\text{ THz}$ retention ceiling. Simply down-weighting the newly acquired data to $w = 0.168$ fails to restore compliance, leaving phonon drift at $0.0587\text{ THz}$. Conversely, augmenting the equal-weight objective ($w = 0.500$) with the pairwise relative-energy penalty $\mathcal{L}_{\Delta e}$ suppresses phonon RMS drift by nearly half to $0.0365\text{ THz}$---passing all physical retention guardrails---while preserving a substantial audit force reduction ($43.11~\mevA$). Combining reduced new-data weighting ($w = 0.168$) with $\mathcal{L}_{\Delta e}$ further lowers phonon drift to $0.0229\text{ THz}$ at the expense of target responsiveness ($44.14~\mevA$). The relative-energy objective at $w = 0.500$ therefore provides the best accuracy--retention trade-off among the tested retention-compliant settings and serves as the standard learner for our multi-round benchmark.

\begin{table}[tbp]
\centering\scriptsize
\caption{\textbf{Matched objective ablation on eight MD-derived training configurations.} All runs share an identical initialization (QP17), random seed and 100-update budget. Audit force errors are evaluated on six held-out frames from independent MD trajectories. Retention status reflects conjunctive compliance across historical errors, lattice and elastic constants, and harmonic phonon drift ($0.05\text{ THz}$ RMS limit). All four objectives passed retention checks on the separate eight-frame candidate-pool subset.}
\label{tab:ablation}
\setlength{\tabcolsep}{3pt}
\renewcommand{\arraystretch}{1.15}
\begin{tabularx}{\textwidth}{@{}>{\raggedright\arraybackslash}Xcccc@{}}
\toprule
Objective formulation & \shortstack{New-data\\weight ($w$)} & \shortstack{Audit force RMSE\\($\mevA$)} & \shortstack{Phonon RMS\\drift (THz)} & \shortstack{Retention\\status}\\
\midrule
Original base loss & 0.500 & 41.36 & 0.0645 & Fail\\
Cardinality-reduced weight & 0.168 & 41.82 & 0.0587 & Fail\\
Relative-energy anchored ($\mathcal{L}_{\Delta e}$) & 0.500 & 43.11 & 0.0365 & Pass\\
Reduced weight + $\mathcal{L}_{\Delta e}$ & 0.168 & 44.14 & 0.0229 & Pass\\
\bottomrule
\end{tabularx}
\end{table}

The physical basis for an energy-difference penalty stabilizing second-order vibrational spectra is illustrated by a local harmonic expansion of the potential-energy surface. For a small atomic displacement field $u$ relative to a mechanical equilibrium reference structure, the configurational energy difference satisfies
\begin{equation}
E(u) - E(0) = \frac{1}{2}u^\top K u + \mathcal{O}(\|u\|^3), 
\label{eq:curvature}
\end{equation}

where $K$ is the harmonic force-constant matrix governing phonon eigenfrequencies. Because the historical replay pool includes strained bulk configurations, enforcing tight pairwise energy differences across compatible historical groups ($s_{\Delta e} = 1~\mevat$, compared with $s_E = 20~\mevat$ for absolute energies) constrains the sampled energy landscape, including its quadratic response near equilibrium. Protocol-compatible relative-energy anchoring thus supplies an implicit constraint on sampled curvature, stabilizing harmonic lattice dynamics without requiring explicit second-derivative supervision during backpropagation.

\subsection{Retention-constrained acquisition improves defect-domain accuracy without breaching bulk-mechanical tolerances}

\begin{figure}[tbp]
\centering\includegraphics[width=\textwidth]{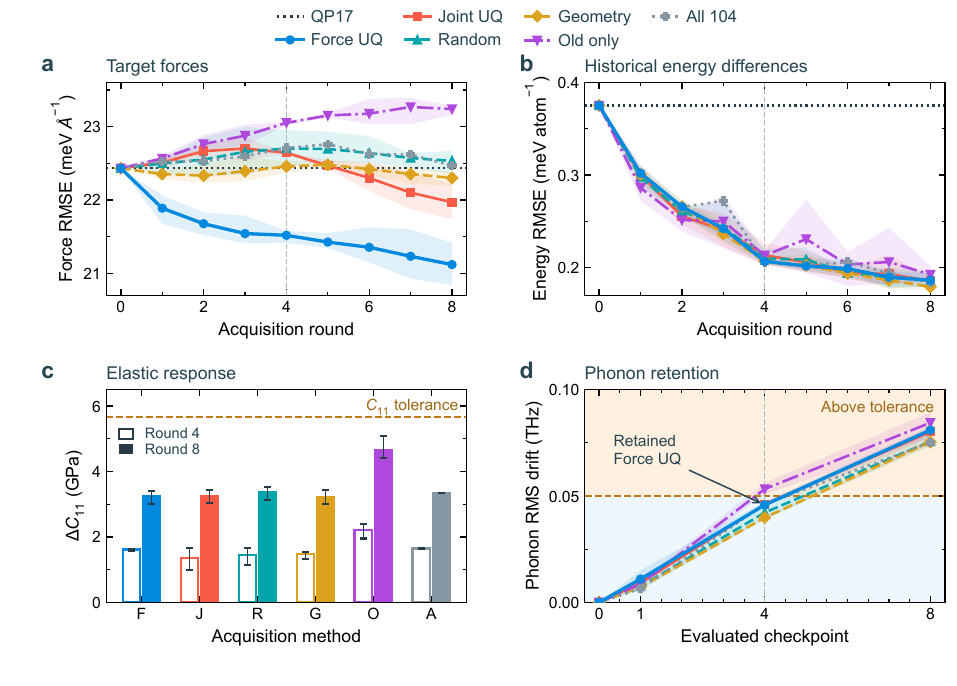}
\caption{\textbf{Multi-round acquisition efficiency and bulk property retention under relative-energy anchoring.} \textbf{a}, Development-set target force RMSE and \textbf{b}, historical replay relative-energy RMSE across eight acquisition rounds starting from the common QP17 baseline at round 0. \textbf{c}, Signed shift in elastic constant $C_{11}$ ($\Delta C_{11}$) at rounds 4 (open bars) and 8 (filled bars) relative to the $5\%$ elastic tolerance (dashed horizontal line). \textbf{d}, Harmonic phonon RMS drift evaluated at rounds 0, 1, 4 and 8, highlighting the retention-compliant zone ($\le 0.05\text{ THz}$, blue shading) and tolerance-violation zone ($> 0.05\text{ THz}$, orange shading). Vertical dashed lines in \textbf{a}, \textbf{b} and \textbf{d} mark the retained Force-UQ operating point at round 4. Curves and bars denote three-seed means; shaded bands and error bars indicate seed minimum--maximum ranges (\texttt{All 104} uses a single seed). Method abbreviations in \textbf{c}: F, Force UQ; J, Joint UQ; R, Random; G, Geometry; O, Old only; A, All 104.}
\label{fig:learning}
\end{figure}

Having established that anchoring is necessary, we ask whether uncertainty-guided acquisition converts that safety into accuracy gains under a fixed update budget. Across the eight-round fixed-pool benchmark under the relative-energy learner, Force UQ consistently achieves lower target force errors than all competing sampling strategies (Fig.~\ref{fig:learning}a). Its mean development force RMSE drops sharply from $22.43~\mevA$ at round 0 to $21.89~\mevA$ after round 1, reaching $21.52~\mevA$ at round 4 using 32 acquired labels and 400 gradient updates. At the same four-round budget, uniform random sampling and geometry-diversity selection yield $22.70$ and $22.46~\mevA$, respectively. Notably, the entire three-seed range of Force UQ at round 4 ($21.42\text{--}21.58~\mevA$) lies strictly below the seed ranges of both random ($22.46\text{--}22.95~\mevA$) and geometric ($22.37\text{--}22.53~\mevA$) controls, showing a consistent acquisition advantage across the three paired runs.

Comparing Force UQ against the full-pool reference (\texttt{All 104}) reveals why selective data acquisition outperforms brute-force data volume under a finite optimization budget. Training uniformly on all 104 candidate labels from the first round yields a development force RMSE of $22.71~\mevA$ after 400 updates---substantially worse than the $21.52~\mevA$ achieved by Force UQ using fewer than one-third of the labels (Table~\ref{tab:checkpoint}). Active acquisition reshapes not only label availability but also the stochastic gradient distribution. Because the 104-configuration pool spans perturbed bulk, planar-slip and grain-boundary environments, uniform sampling assigns each high-strain $\Sigma 5$ grain-boundary configuration a draw probability of only $1/104$ per new-data draw. By contrast, Force UQ isolates eight loaded $\Sigma 5$ configurations in round 1, assigning them an initial draw probability of $1/8$ and retaining them within the growing cumulative pool for repeated reinforcement across subsequent rounds.

This targeted curriculum emerges naturally from the spectral structure of the acquisition variance. Expressing the eigendecomposition of the conditional readout covariance as $Cv_k = c_k v_k$, Eq.~\eqref{eq:uq} expands as
\begin{equation}
u_F^2(X) = \frac{1}{3N}\sum_{k} c_k \|J_F(X)v_k\|_2^2. 
\label{eq:spectral}
\end{equation}

A candidate configuration receives a high acquisition score when its force sensitivities project strongly ($\|J_F(X)v_k\|_2^2$) along readout parameter directions $v_k$ that remain weakly constrained by the historical prior and previously acquired data (large $c_k$). Sequential precision updates within each batch immediately damp $c_k$ along directions spanned by selected structures, while full-parameter training between rounds changes the force Jacobian $J_F(X)$ to reassess the remaining sensitivity directions. Task-aware sensitivity projection thus provides task-specific acquisition information beyond unsupervised geometric space-filling.

At the scheduled four-round operating point, the target force gains of Force UQ coexist with complete compliance across all bulk retention guardrails (Table~\ref{tab:checkpoint} and Fig.~\ref{fig:learning}b--d). Historical relative-energy RMSE improves markedly from $0.375$ to $0.207~\mevat$ (Fig.~\ref{fig:learning}b), while replay and independent historical-holdout force errors increase by only $0.50$ and $0.42~\mevA$, respectively. Mean equilibrium lattice constant drift remains below $0.01\%$, shifts across the independent elastic constants ($C_{11}, C_{12}, C_{44}$) remain below $1.5\%$ (SI; $C_{11}$ shown in Fig.~\ref{fig:learning}c), and phonon RMS drift spans $0.0449\text{--}0.0474\text{ THz}$ (mean $0.0458\text{ THz}$), safely beneath the $0.05\text{ THz}$ tolerance (Fig.~\ref{fig:learning}d; full tabular metrics in the SI).

\begin{table}[tbp]
\centering\scriptsize
\caption{\textbf{Target accuracy and baseline property retention at the round-4 operating point.} Values represent means across three paired seeds (except QP17 and the single-seed \texttt{All 104} reference). Force RMSEs are reported in $\mevA$, historical relative-energy RMSEs in $\mevat$, and harmonic phonon RMS drift in THz. The final column records the fraction of seed runs satisfying every predefined retention guardrail.}
\label{tab:checkpoint}
\setlength{\tabcolsep}{3pt}
\renewcommand{\arraystretch}{1.15}
\begin{tabularx}{\textwidth}{@{}>{\raggedright\arraybackslash}Xcccccc@{}}
\toprule
Acquisition method & \shortstack{Acquired\\labels} & \shortstack{Target\\force RMSE} & \shortstack{Replay\\force RMSE} & \shortstack{Replay\\$\Delta E$ RMSE} & \shortstack{Phonon\\RMS drift} & \shortstack{Retention\\pass rate}\\
\midrule
QP17 (Baseline) & 0 & 22.432 & 8.478 & 0.375 & 0.0000 & ---\\
Force UQ & 32 & 21.518 & 8.980 & 0.207 & 0.0458 & 3/3\\
Joint UQ & 32 & 22.643 & 9.043 & 0.213 & 0.0462 & 3/3\\
Random & 32 & 22.703 & 8.998 & 0.210 & 0.0422 & 3/3\\
Geometry & 32 & 22.457 & 8.986 & 0.207 & 0.0399 & 3/3\\
Old only (Replay control) & 0 & 23.052 & 8.827 & 0.213 & 0.0533 & 0/3\\
All 104 (Full pool) & 104 & 22.710 & 9.008 & 0.206 & 0.0462 & 1/1\\
\bottomrule
\end{tabularx}
\end{table}

Crucially, the \texttt{Old only} control reveals a counter-intuitive phenomenon: training exclusively on historical replay data without introducing new defect configurations does not guarantee baseline property preservation. By round 4, all three \texttt{Old only} runs violate the phonon retention threshold (mean drift $0.0533\text{ THz}$; Table~\ref{tab:checkpoint}), and by round 8 this branch exhibits the largest elastic constant distortion ($\Delta C_{11} \approx 4.7\text{ GPa}$, compared with $\sim 3.3\text{ GPa}$ across all defect-acquiring branches; Fig.~\ref{fig:learning}c) alongside degraded target force accuracy ($23.24~\mevA$ at round 8). Because the 79-configuration replay buffer represents only a compact subsample of the original QP17 training distribution, prolonged optimization solely on this restricted subset shifts the sampled fitting problem away from the original training distribution. The smaller elastic and harmonic changes in the defect-acquiring branches are consistent with broader geometric constraints mitigating this drift.

Extending optimization to round 8 (64 acquired labels, 800 updates) drives Force-UQ target force error down further to $21.12~\mevA$, but pushes phonon RMS drift to $\sim 0.08\text{ THz}$, breaching the harmonic tolerance across every branch (Fig.~\ref{fig:learning}a,d). Periodic physical-property auditing thus identifies round 4 as the latest retention-compliant operating point among the scheduled complete evaluations, retaining the defect accuracy gain before the round-8 harmonic drift exceeds physical tolerances.

\subsection{Gains transfer to plastic-deformation observables: defect energies, deviatoric stresses and slip-path energetics}

\begin{figure}[tbp]
\centering\includegraphics[width=\textwidth]{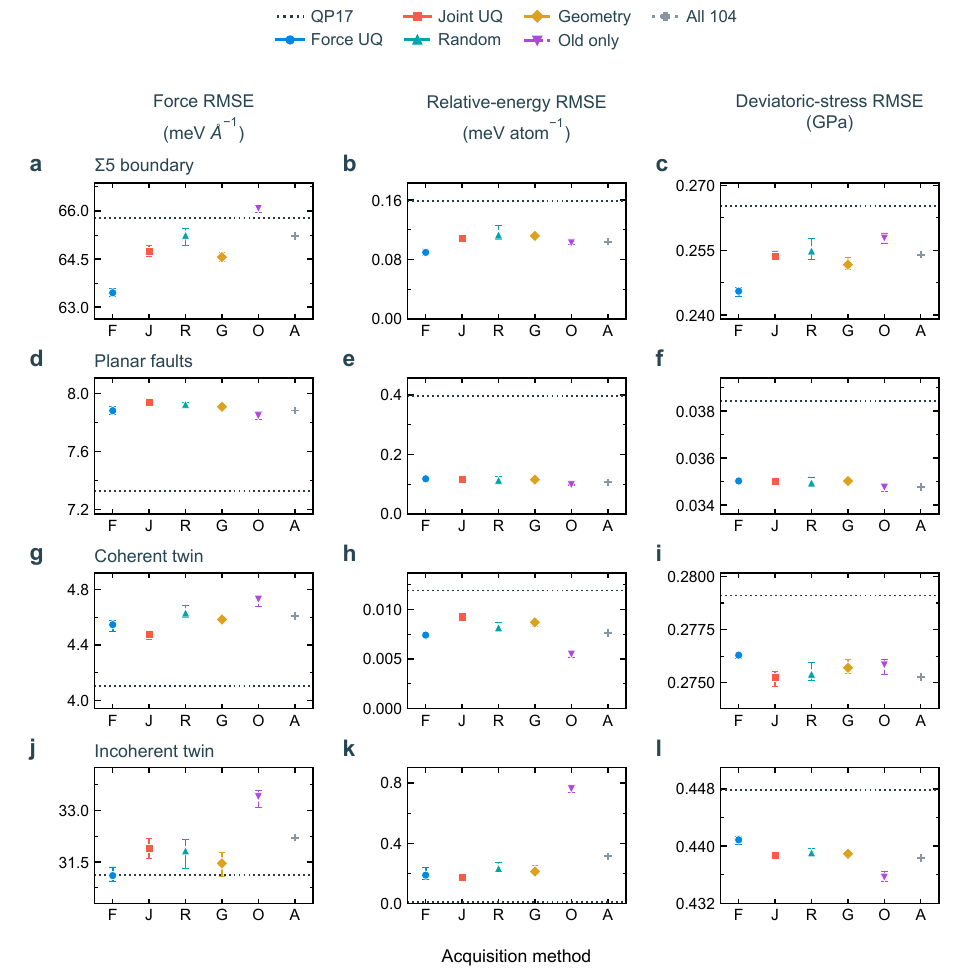}
\caption{\textbf{Transfer to plastic-deformation observables across the external archive at the round-4 operating point.} Rows resolve four distinct defect families: mechanically loaded $\Sigma 5$ grain boundaries (\textbf{a--c}, two states), $(111)$ planar stacking faults (\textbf{d--f}, three states), perturbed coherent twin boundaries (\textbf{g--i}, one state) and incoherent twin boundaries (\textbf{j--l}, two states). Columns report out-of-sample force RMSE (\textbf{a, d, g, j}), within-family relative-energy RMSE (\textbf{b, e, h, k}) and deviatoric-stress RMSE (\textbf{c, f, i, l}). Symbols and error bars denote three-seed means and minimum--maximum ranges; horizontal dotted lines mark the QP17 baseline (\texttt{All 104} uses a single seed; method abbreviations follow Fig.~\ref{fig:learning}).}
\label{fig:external}
\end{figure}

Evaluating the round-4 checkpoints on the external defect archive tests the central claim of this work: that the adapted potential improves plastic-deformation observables, defect energetics, deviatoric stresses and slip-path energetics, and not merely the development force metric (Fig.~\ref{fig:external}). Across the archive, Force UQ slashes relative-energy RMSE in planar stacking faults by $70\%$ (from $0.398$ to $0.118~\mevat$; Fig.~\ref{fig:external}e) and in $\Sigma 5$ grain boundaries by $43\%$ (from $0.159$ to $0.090~\mevat$; Fig.~\ref{fig:external}b), while also reducing perturbed coherent-twin energy error (Fig.~\ref{fig:external}h). Remarkably, even though virial stresses are never fitted during training, Force UQ lowers deviatoric-stress RMSE across all four defect families (Fig.~\ref{fig:external}c,f,i,l). In the high-strain $\Sigma 5$ boundary family, Force UQ reduces reference force RMSE from $65.77$ to $63.45~\mevA$, outperforming both random ($65.25~\mevA$) and geometric ($64.56~\mevA$) baselines (Fig.~\ref{fig:external}a). In low-error near-symmetric families (planar faults and coherent twins, where baseline force errors are already $4\text{--}8~\mevA$), small increases in force RMSE of $\sim 0.4\text{--}0.5~\mevA$ accompany the substantial improvements in stacking-fault energies and stresses (Fig.~\ref{fig:external}d,g), while incoherent-twin relative-energy error rises from $0.010$ to $0.188~\mevat$ (Fig.~\ref{fig:external}k). Presenting energy, force and stress metrics jointly across defect topologies thus exposes both the primary transfer gains and the localized trade-offs of adaptation.

Comparing Force UQ with Joint UQ illuminates how force Jacobians ($J_F$) and deviatoric-stress Jacobians ($J_D$) probe distinct physical regimes of the potential-energy surface. Whereas forces measure energy derivatives with respect to atomic displacements, deviatoric stresses measure collective energy variations under homogeneous volume-preserving cell strain. Consequently, during round-1 acquisition, Force UQ prioritizes mechanically loaded $\Sigma 5$ grain-boundary configurations, whereas Joint UQ prioritizes perturbed FCC/HCP bulk cells. Although localized planar slip is kinematically distinct from homogeneous cell strain, both deformations interrogate shear-coupled readout directions.

\begin{figure}[tbp]
\centering\includegraphics[width=\textwidth]{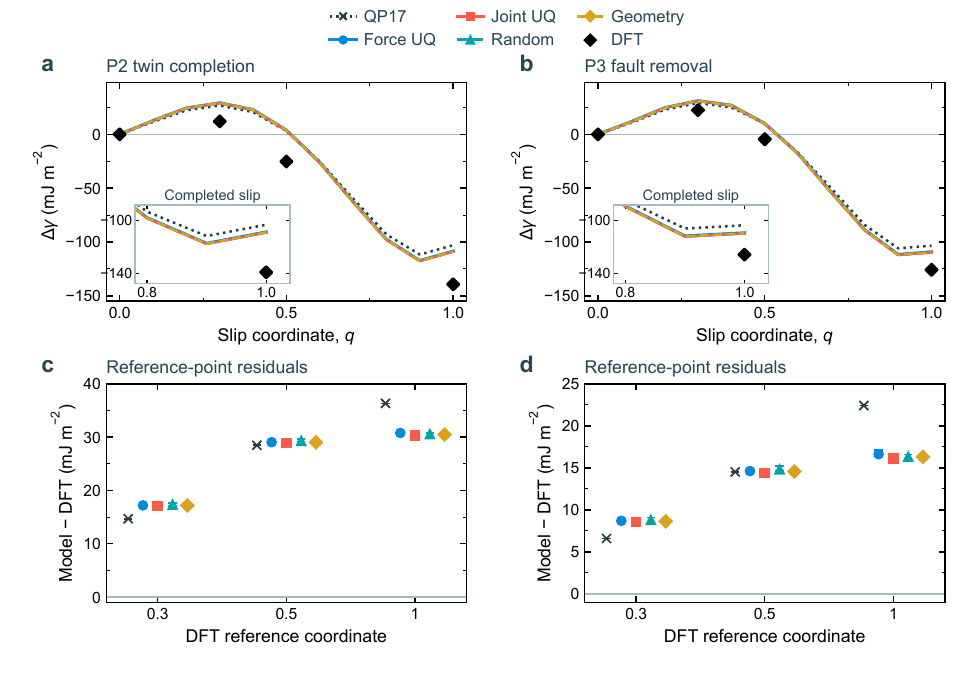}
\caption{\textbf{Planar-slip energy landscapes and reference-point residuals at round 4.} \textbf{a, b}, Fault energy profiles ($\Delta\gamma$, referenced to each model's relaxed \texttt{101} precursor) along prescribed shear coordinates $q$ for route P2 (\texttt{101} $\rightarrow$ \texttt{111} twin completion, \textbf{a}) and route P3 (\texttt{101} $\rightarrow$ \texttt{100} intrinsic fault restoration, \textbf{b}). Insets magnify the completed-slip regime ($q \in [0.8, 1.0]$); black diamonds denote discrete DFT reference coordinates. \textbf{c, d}, Signed energy residuals ($\Delta\gamma_{\mathrm{model}} - \Delta\gamma_{\mathrm{DFT}}$) evaluated at the three non-zero DFT reference coordinates ($q = 0.3, 0.5, 1.0$) along P2 (\textbf{c}) and P3 (\textbf{d}). Curves and symbols represent three-seed means with minimum--maximum seed ranges.}
\label{fig:paths}
\end{figure}

This mechanistic distinction manifests along the constrained planar-slip pathways P2 (\texttt{101} $\rightarrow$ \texttt{111} twin completion) and P3 (\texttt{101} $\rightarrow$ \texttt{100} fault removal) at round 4 (Fig.~\ref{fig:paths}). While Force UQ reduces the P2/P3 reference-point energy RMSEs from the QP17 baseline of $27.97 / 15.89~\mjm$ to $26.38 / 13.74~\mjm$, Joint UQ achieves slightly lower pathway errors of $26.11 / 13.42~\mjm$ (compared with $26.42 / 13.72~\mjm$ for random and $26.26 / 13.58~\mjm$ for geometric sampling). Inspecting the full slip profiles (Fig.~\ref{fig:paths}a,b) and signed reference-point residuals (Fig.~\ref{fig:paths}c,d) reveals that error reduction is concentrated primarily at the completed-slip endpoints ($q = 1.0$), where model-minus-DFT residuals drop by $\sim 6~\mjm$, whereas intermediate reference-point residuals ($q = 0.3, 0.5$) change less and can increase. Incorporating deviatoric-stress sensitivity into acquisition thus modestly shifts the response toward the tested slip observables, while pure force uncertainty gives the stronger development-force improvement.

Comparing slip-path evolution across training objectives further clarifies the role of historical energy anchoring. When unconstrained by $\mathcal{L}_{\Delta e}$, the original-objective Joint-UQ branch drives P2/P3 RMSEs down aggressively to $18.95 / 6.89~\mjm$ by round 8 (versus $21.50 / 9.49~\mjm$ for random acquisition), but at the cost of breaching the bulk phonon retention limit (complete original-objective curves in the SI). Relative-energy anchoring deliberately restrains the magnitude of this pathway shift to preserve the tested harmonic response at the round-4 operating point. Notably, neither objective reverses QP17's near-degenerate ordering of the completed twin and intrinsic fault endpoints ($\gamma_{\mathrm{P3}} - \gamma_{\mathrm{P2}} \approx -0.4\text{ to }-0.8~\mjm$, versus $+13.50~\mjm$ in DFT). Because fine-tuning lowers both endpoint energies by a comparable common offset (Fig.~\ref{fig:paths}a,b insets), absolute pathway RMSE improves without generating the differential endpoint-energy correction required to split their near-degeneracy. Resolving subtle stacking-fault polymorphism thus represents a distinct physical target beyond small-sample force-driven adaptation.

\subsection{Closed-loop acceptance gating separates informative acquisition from safe deployment}

Finally, the trajectory-driven MD--DFT pilot extends our workflow from a static candidate pool to on-the-fly configuration generation. Propagating QP17 across $300\text{ K}$ bulk FCC and $600\text{ K}$ thin-twin supercells yields stable dynamics without triggering a threefold relative-uncertainty excursion; nevertheless, the prescribed exploration quota extracts four candidate snapshots per branch for first-principles labelling. Updating QP17 on these four frames under the original objective reduces the four-frame independent-trajectory monitoring force RMSE from $49.72~\mevA$ to $44.24~\mevA$ for Joint UQ, outperforming the random-selection update ($44.77~\mevA$). Despite this target accuracy gain, the automated physical-property gate rejects both candidate updates because their historical elastic-group energy errors and phonon RMS drifts exceed the predefined guardrails, retaining the baseline QP17 potential for subsequent dynamics (Fig.~\ref{fig:workflow}c).

This closed-loop test validates a critical architectural principle for autonomous active learning: decoupling informative data acquisition from model deployment. While relative trajectory uncertainty can remain quiescent during short thermal sampling, independent DFT auditing still detects substantial absolute force errors ($> 49~\mevA$) that benefit from targeted retraining. Conversely, acquiring informative frames and reducing validation force error does not guarantee that a gradient update is physically safe to deploy. As demonstrated in our matched ablation (Table~\ref{tab:ablation}), replacing the original loss with the relative-energy-anchored learner ($\mathcal{L}_{\Delta e}$) allows the combined MD-derived labels to be assimilated within the $0.05\text{ THz}$ phonon guardrail. Pairing uncertainty-guided exploration with curvature-anchoring regularization and strict physical acceptance gating thus provides a safeguard against measured baseline degradation in potential refinement. This closed-loop test validates the architectural principle behind retention-constrained fine-tuning: informative acquisition and safe deployment are separate decisions, and only the second determines whether an adapted potential may be used for non-equilibrium dynamics.

\section{Conclusions}
Selective adaptation of a pre-trained interatomic potential requires balancing non-equilibrium defect accuracy against bulk-mechanical retention. By uniting conditional linear-readout uncertainty with protocol-compatible relative-energy anchoring and conjunctive bulk-mechanical gating, we have demonstrated a reproducible, retention-constrained workflow for adapting the aluminium MACE potential QP17 to non-equilibrium defect configurations while keeping the tested bulk response within predefined tolerances. Across multi-round fixed-pool benchmarks, force-uncertainty acquisition concentrates limited gradient updates onto high-strain grain-boundary environments, achieving lower target force errors with 32 labels than random sampling, geometric diversity or uniform training across all 104 pool configurations. At the four-round operating horizon, all force-uncertainty models satisfy strict lattice, elastic and harmonic phonon retention bounds while simultaneously reducing historical relative-energy errors.

Our component analyses clarify how retention-constrained fine-tuning achieves the plasticity--stability balance: (1) Non-equilibrium defect accuracy is improved by projecting readout sensitivities onto under-constrained parameter directions and updating precision matrices sequentially within each batch, so that limited gradient updates concentrate on the far-from-equilibrium configurations that govern plastic response. (2) Bulk-mechanical retention is protected by protocol-compatible relative-energy anchoring, which constrains pairwise energy differences within historical subsets and thereby implicitly regularizes the near-equilibrium force-constant curvature that sets elastic and vibrational response, without explicit phonon supervision. (3) Deployment is gated by a conjunctive bulk-mechanical audit, ensuring that gains in defect forces, stacking-fault energies and deviatoric stresses are retained only when lattice, elastic and vibrational tolerances are respected; the closed-loop MD--DFT pilot shows that this gate can reject an accuracy-improving update. As pre-trained and foundation MLIPs become the standard starting point for atomistic materials modelling, this modular strategy---targeted sensitivity acquisition, curvature-anchoring replay and automated physical acceptance gating---offers a practical blueprint for domain-specific fine-tuning with verifiable baseline-retention criteria.

\section*{Data and code availability}
The configurations, reference labels, acquisition histories, model checkpoints and analysis scripts underlying this study have been retained. Repository and accession information will be inserted in the submission version.
\section*{Author contributions, acknowledgements and declarations}
Author names and affiliations, contributions, funding and competing-interest declarations are to be completed by the authors before submission.

\end{document}